%Paper: hep-th/9511154
%From: Philip Argyres <argyres@physics.rutgers.edu>
%Date: Wed, 22 Nov 1995 11:31:41 -0500

%%%%%%%%%%%%%%%%%%%%%%%%%%%%%%%%%%%%%%%%
\newif\ifbbB\bbBfalse                %%%
%%%       BLACKBOARD BOLD FONT       %%%
%%% Comment-out the next line if you %%%
%%% do NOT have the blackboard bold  %%%
%%% (mssb) fonts:                    %%%
\bbBtrue                             %%%
%%%%%%%%%%%%%%%%%%%%%%%%%%%%%%%%%%%%%%%%

%%%%%%%%%%%%%%%%%%%%%%%%%%%%%%%%%%%%%%%%%%%%%%%%%%%%%%%%%%%%%%%%%
%                                               		%
% NEW N=2 SUPERCONFORMAL FIELD THEORIES IN FOUR DIMENSIONS	%
%                                               		%
% By  P.C. Argyres, M.R. Plesser, N. Seiberg, and E. Witten	%
%                                               		%
% Contact  argyres@physics.rutgers.edu           		%
%   	   (908) 445-5290                         		%
%  	   Department of Physics and Astronomy  		%
%  	   Rutgers University                   		%
%  	   Piscataway NJ 08855, USA              		%
%                                               		%
%%%%%%%%%%%%%%%%%%%%%%%%%%%%%%%%%%%%%%%%%%%%%%%%%%%%%%%%%%%%%%%%%

\input harvmac
\overfullrule=0pt
\def\Title#1#2{\rightline{#1}
 \ifx\answ\bigans
  \nopagenumbers\pageno0\vskip1in\baselineskip15pt plus1pt minus1pt
 \else
  \def\listrefs{\footatend\vskip 1in\immediate\closeout\rfile\writestoppt
   \baselineskip=14pt\centerline{{\bf References}}\bigskip{\frenchspacing
   \parindent=20pt\escapechar=` \input refs.tmp\vfill\eject}\nonfrenchspacing}
  \pageno1\vskip.8in\fi
\centerline{\titlefont #2}\vskip .5in}

\ifx\answ\bigans
 \def\tcbreak#1{}
\else
 \def\tcbreak#1{\cr&{#1}}
\fi
\ifbbB
 \message{If you do not have msbm (blackboard bold) fonts,}
 \message{change the option at the top of the tex file.}
 \font\blackboard=msbm10 % scaled \magstep1
 \font\blackboards=msbm7 \font\blackboardss=msbm5
 \newfam\black \textfont\black=\blackboard
 \scriptfont\black=\blackboards \scriptscriptfont\black=\blackboardss
 \def\Bbb#1{{\fam\black\relax#1}}
\else
 \def\Bbb{\bf}
\fi

\def\NPB#1#2#3{{\sl Nucl. Phys.} \underbar{B#1} (#2) #3}
\def\PLB#1#2#3{{\sl Phys. Lett.} \underbar{#1B} (#2) #3}
\def\PRL#1#2#3{{\sl Phys. Rev. Lett.} \underbar{#1} (#2) #3}
\def\PRD#1#2#3{{\sl Phys. Rev.} \underbar{D#1} (#2) #3}
\def\CMP#1#2#3{{\sl Comm. Math. Phys.} \underbar{#1} (#2) #3}
\def\til{\widetilde}
\def\bar{\overline}
\def\vev#1{{\langle #1\rangle}}
\def\bC{{\Bbb C}}
\def\bZ{{\Bbb Z}}
\def\CN{{\cal N}}
\def\tu{{\tilde u}}
\def\tj{{\tilde\jmath}}

\Title{RU-95-81, WIS-95-59, IASSNS-HEP-95/95}
{\vbox{\centerline{New $\CN{=}2$ Superconformal Field Theories}
\medskip
\centerline{in Four Dimensions}}}
\centerline{Philip C. Argyres$^1$, M. Ronen Plesser$^2$,
Nathan Seiberg$^1$, and Edward Witten$^3$}
\smallskip
\centerline{{}$^1$\it Department of Physics and Astronomy, Rutgers
University, Piscataway NJ 08855, USA}
\centerline{{}$^2$\it Department of Particle Physics, Weizmann
Institute of Science, 76100 Rehovot Israel}
\centerline{{}$^3$\it School of Natural Sciences, Institute for
Advanced Study, Princeton NJ 08540, USA}
\bigskip
\noindent
{\bf Abstract:} New examples of $\CN{=}2$ supersymmetric conformal field
theories are found as fixed points of $SU(2)$ $\CN{=}2$ supersymmetric
QCD.  Relations among the scaling dimensions of their relevant chiral
operators, global symmetries, and Higgs branches are understood
in terms of the general structure of relevant
deformations of non-trivial $\CN{=}2$ conformal field theories.
The spectrum of scaling dimensions found are all those compatible
with relevant deformations of a $y^2 = x^3$ singular curve.
\Date{November 1995}
%\draftmode

\nref\Niv{M. Sohnius and P. West, ``Conformal Invariance in N=4
	Supersymmetric Yang-Mills Theory,'' \PLB{100}{1981}{245}.}
\nref\fNii{P. Howe, K. Stelle, and P. West, ``A Class of Finite
	Four-Dimensional Supersymmetric Field Theories,''
	\PLB{124}{1983}{55}.}
\nref\fNi{A. Parkes and P. West, \PLB{138}{1984}{99}; P. West,
	\PLB{137}{1984}{371}; D.R.T. Jones and L. Mezincescu,
	\PLB{138}{1984}{293}; S. Hamidi, J. Patera, and J. Schwarz,
	\PLB{141}{1984}{349}; S. Hamidi and J. Schwarz, \PLB{147}{1984}{301};
	W. Lucha and H.  Neufeld, \PLB{174}{1986}{186}, \PRD{34}{1986}{1089};
	D.R.T. Jones, \NPB{277}{1986}{153}; A.V. Ermushev, D.I. Kazakov and
	O.V. Tarasov, \NPB{281}{1987}{72}; X.-D. Jiang and X.-J. Zhou,
	\PRD{42}{1990}{2109}; D.I. Kazakov, {\sl Mod. Phys. Lett.}
	{\underbar A2} (1987) {663}, {\sl Ninth Dubna Conf. on the Problems
	of Quantum Field Theory}, Dubna, 1990; O. Piguet and K. Sibold,
	{\sl Int. J. Mod. Phys.} {\underbar A1} (1986) {913},
	\PLB{177}{1986}{373}; C. Lucchesi, O. Piguet, and K. Sibold,
	{\sl Conf. on Differential Geometrical Methods in Theoretical
	Physics}, Como, 1987, {\sl Helv. Phys. Acta} {\underbar 61} (1988)
	{321}; R.G. Leigh and M.J. Strassler, hep-th/9503121,
	\NPB{477}{1995}{95}.}
\nref\Noqcd{T. Banks and A. Zaks, ``On the Phase Structure of Vectorlike
	Gauge Theories with Massless Fermions,'' \NPB{196}{1982}{189}.}
\nref\Sei{N. Seiberg, ``Electric-Magnetic Duality in Supersymmetric
        non-Abelian Gauge Theories,'' hep-th/9411149, \NPB{435}{1995}{129}.}
\nref\AD{P.C. Argyres and M.R. Douglas, ``New Phenomena in SU(3)
        Supersymmetric Gauge Theory,'' hep-th/9505062, \NPB{448}{1995}{93}.}
\nref\SWII{N. Seiberg and E. Witten, ``Monopoles, Duality, and Chiral
        Symmetry Breaking in N=2 Supersymmetric QCD,'' hep-th/9408099,
        \NPB{431}{1994}{484}.}
\nref\Mack{G. Mack, ``All Unitary Ray Representations of the Conformal
	Group SU(2,2) with Positive Energy,'' \CMP{55}{1977}{1}.}
\nref\Adler{S.L. Adler, ``Short-Distance Behavior of Quantum
	Electrodynamics and an Eigenvalue Condition for $\alpha$,''
	\PRD{5}{1972}{3021}.}
\nref\DP{V.K. Dobrev and V.B. Petkova, ``All Positive Energy Unitary
	Irreducible Representations of Extended Conformal Supersymmetry,''
	\PLB{162}{1985}{127}.}
\nref\APSe{P.C. Argyres, M.R. Plesser, and N. Seiberg, to appear.}
\nref\dWLVP{B. de Wit, P.G. Lauwers, and A. Van Proeyen, ``Lagrangians of
        N=2 Supergravity-Matter Systems,'' \NPB{255}{1985}{569}.}
\nref\SWI{N. Seiberg and E. Witten, ``Electric-Magnetic Duality,
        Monopole Condensation, and Confinement in N=2 Supersymmetric
        Yang-Mills Theory,'' hep-th/9407087, \NPB{426}{1994}{19}.}
\nref\DW{R. Donagi and E. Witten, ``Supersymmetric Yang-Mills Theory and
	Integrable Systems,'' hep-th/9510101.}
\nref\APS{P.C. Argyres, M.R. Plesser, and A.D. Shapere, ``The Coulomb
        Phase of N=2 Supersymmetric QCD,'' hep-th/9505100,
        \PRL{75}{1995}{1699}.}

\newsec{Introduction}

The study of conformal field theory (CFT) is an important starting
point for the study of field theory.  CFT's describe the IR behavior of
asymptotically free theories and the behavior at all scales of scale
invariant theories.  The best known examples of non-trivial four
dimensional CFT's are based on finite supersymmetric gauge theories
\refs{\Niv - \fNi}.  These are scale invariant theories labeled by
their coupling constant $\tau$, a truly marginal operator which does
not run.  Their short distance behavior is not free.  Other examples
occur in asymptotically free theories whose IR behavior may be a
non-trivial CFT or a free CFT in terms of new degrees of freedom.
Examples of non-trivial CFT's {}from asymptotically free theories are
QCD with many flavors \Noqcd, $\CN{=}1$ supersymmetric QCD with many
flavors \Sei, and the fixed points of pure $\CN{=}2$ gauge theory
studied in \AD.  In this letter we find new examples of $\CN{=}2$ CFT's
as fixed points of $\CN{=}2$ $SU(2)$ QCD, using the exact solution of
\SWII, and systematize some of their properties using superconformal
invariance.

In section 2 we discuss the basic conditions on non-trivial fixed
points coming {}from conformal invariance.  One consequence of these
conditions is that there can be no non-trivial point with vector fields
unless there are nonvanishing electric and magnetic currents in
the theory.  We also give a general picture of the structure of
non-trivial $\CN{=}2$ CFT's in terms of their relevant deformations.
In section 3 we catalog the non-trivial CFT's in $\CN{=}2$ $SU(2)$ QCD
by determining scaling dimensions of chiral fields, their global
symmetries, and Higgs branches.  Four inequivalent nontrivial CFT's are
found, which can be characterized as vacua where a monopole and $N_f$
quarks simultaneously become massless, for $N_f = 1,2,3,4$.  The
$N_f{=}1$ CFT is equivalent to the fixed point in $SU(3)$ Yang-Mills
theory studied in \AD.  The $N_f{=}4$ CFT is the scale-invariant $SU(2)$
4-flavor theory studied in \SWII.  The scaling dimensions of relevant
chiral operators of all four theories are explained in terms of the
relevant deformations of a singular curve $y^2{=}x^3$.

\newsec{Conditions {}from Conformal Invariance}

We start by discussing four-dimensional CFT's with no supersymmetry.
Fields are in representations of the conformal algebra, labeled by their
scaling dimension $D$ and their $SU(2) {\times} SU(2)$ Lorentz spins
$(j,\tj)$. There is a one to one map between local operators and the
states they create by acting on the conformally invariant vacuum.
Primary states are those annihilated by the special conformal
generators; descendant fields are created by acting with momentum
generators on primary states.  {}From the representation theory of the
conformal algebra \Mack, unitary ``chiral'' primary fields, those with
either $j$ or $\tj=0$, satisfy the inequality $D \ge j {+} \tj {+} 1$,
with equality only for free fields.  Non-chiral primary fields satisfy
$D \ge j {+} \tj {+} 2$.

For example, consider a field strength operator $F_{\mu\nu}$ of
conformal dimension $D$, which is the sum of two conformal primary
fields $F^\pm = F \pm {}^*F$ with spins $(1,0)$, $(0,1)$.  In the
conformal algebra the states associated with the conserved currents
$J^\pm_\mu = \partial^\nu F_{\mu\nu}^\pm = {}^*dF^\pm$ satisfy
$||J^\pm_\mu \rangle |^2 = 2(D{-}2)$.  We see that unitarity and the
conformal algebra imply $D{\ge}2$.  Equality holds if and only if
$J^\pm=0$, implying the Bianchi identity and free equations of motion
$dF^+ \pm dF^- =0$ (free Maxwell theory).

If $F$ is not free, its dimension is larger than 2 and both $J^+$ and
$J^-$ are not zero.  Since they are descendants of different primary
fields ($F^+$ and $F^-$), they are linearly independent.  Therefore,
both the electric current $J_e \equiv J^+ {+} J^-$ and the magnetic
current $J_m \equiv J^+ {-} J^-$ are non-zero as quantum fields.  We
conclude that in a conformal field theory any interacting field
strength must couple both to electrons and to monopoles.  In
particular, QED without elementary monopoles cannot have a non-trivial
fixed point.\foot{For a related discussion, see \Adler.}

Note that all Abelian gauge charges vanish in a fixed point theory
(though they may still couple to massive degrees of freedom).  In the
case of the interacting $U(1)$ field strength $F$, though we have seen
that its conserved electric and magnetic currents do not vanish, there
is no charge at infinity associated with them, because of the rapid
decay of correlation functions of $F$ due to its anomalous dimension.
This is true even if we include massive or background sources, since
the long-distance behavior of the fields is governed by the conformal
field theory.  If, on the other hand, $F$ were free, then we have seen
that its associated conserved currents, and thus the charges, vanish.
Now, however, massive sources can have long-range fields in this case
since $F$ has its canonical dimension.  (We do not reach a
contradiction by taking the mass of a charged source to zero since its
$U(1)$ couplings flow to zero in the IR.)  Non-Abelian gauge charges need
not vanish in the CFT since the above arguments only apply to
gauge-invariant fields or states.

The $\CN{=}1$ and $\CN{=}2$ superconformal algebras contain the
conformal algebra, and so the above results carry over.  A new feature
of the superconformal algebras is their global $R$-symmetry.  Their
representations are labeled by the $R$-charges as well as the Lorentz
spins and scaling dimension.  For chiral fields the superconformal
algebra implies a relation between the scaling dimension and
$R$-charges.  Since the $R$-symmetry is part of the conformal
algebra,  non-trivial supersymmetric CFT's necessarily carry
non-zero $R$-charges.

The $\CN{=}1$ superconformal algebra includes a $U(1)_R$.  In theories
with an $\CN{=}1$ fixed point, if we can identify the $U(1)_R$ in the
UV, we can determine the dimensions of chiral fields.  Examples of such
fixed points in QCD were first given in \Sei.  Alternatively, the
$U(1)_R$ in the IR could be an accidental symmetry, in which case it
would be difficult to find the dimensions of chiral fields.

The $\CN{=}2$ superconformal algebra includes a $U(1)_R \times
SU(2)_R$.  In QCD the $R$-charges are known in the UV.  If there is no
anomaly in $U(1)_R$, then the theory is finite, there is a marginal
coupling $\tau$, and the dimensions are independent of $\tau$.
Examples are the non-Abelian Coulomb points in the finite $\CN{=}2$
theories \fNii.  On the other hand, if the classical $U(1)_R$ is
anomalous, there may be an accidental $U(1)_R$ in the IR, making it
hard to find dimensions of chiral fields.  Examples of this sort were
given in \AD.

\subsec{Non-trivial $\CN{=}2$ CFT}

In preparation for our exploration of $\CN{=}2$ $SU(2)$ QCD in the
following section, let us examine the $\CN{=}2$ case in more detail.
In addition to the spin and scaling dimension, conformal fields are
labeled by their $U(1)_R$ charge $R$, and their $SU(2)_R$ spin $I$.  A
primary state (annihilated by the superconformal generators) is the
lowest component of a supermultiplet formed by applying the eight
supercharges $Q^\alpha_i$, $\bar Q^{\dot\alpha}_i$ to it.\foot{$\alpha$
and $\dot\alpha$ are Lorentz indices, $i$ is the $SU(2)_R$ index.}
{}{}From the representation theory of the $\CN{=}2$ superconformal algebra
\DP, we learn that chiral ($\tj{=}0$) fields $\phi$ satisfy $\bar
Q_{(i} \phi_{i_1\ldots i_{2I})} {=}0$ and $D = 2I {+} {1\over2}R \ge j
{+} 2I {+} 1$.  A similar relation, given by changing the sign of the
$R$-charge holds for antichiral ($j{=}0$) fields.

The $\CN{=}2$ vector multiplet $U$ has as its lowest component a chiral
primary field $u$ with $I{=}0$ and spin $(0,0)$.  Therefore $D(u) =
{1\over2} R(u) \ge 1$ and the field strength $F^+_{\mu\nu}$ at its
second excited level has $D(F) {\ge} 2$.  When this inequality is
saturated, there is an extra null state at the fourth level giving
$dF^+ {=} 0$.  In $\CN{=}2$ superfield notation, a chiral superfield
$U$,
		\eqn\chirality{
\bar D_{\dot\alpha i}U = 0,
		}
with $R(u){=}2$ satisfies
		\eqn\nullvect{
D^{\alpha(i}D_\alpha^{j)}U = 0.
		}
In this case the field is free, the null state equations for $u$ and
$\bar u$ being equivalent to the vacuum Maxwell equations.  On the
other hand, for an interacting vector multiplet ($D(u) {>} 1$) these
states are no longer null, and we again conclude that the
corresponding electric and magnetic currents cannot vanish as quantum
fields.

$\CN{=}2$ QCD has a moduli space of inequivalent vacua, composed
generically of ``branches'' with some numbers $n_V$ (massless)
$U(1)$ vector multiplets and $n_H$ massless neutral
hypermultiplets.  Branches with $n_H{=}0$ are called Coulomb branches,
with $n_V{=}0$ Higgs branches, and the other cases mixed branches.
These branches may intersect along complex submanifolds, corresponding
to phase transitions.  The Higgs branch is determined classically due
to a nonrenormalization theorem \APSe, which follows from thinking of
the bare masses and the strong-coupling scale $\Lambda$ as the scalar
components of $\CN{=}2$ vector superfields.\foot{This is possible since
one knows how to write explicit Lagrangian with weakly gauged $U(1)$'s
whose scalar part appears as the bare mass or $\Lambda$.} In the
general $\CN{=}2$ effective action \dWLVP\ it is found that the scalar
components of vector multiplets do not appear in the hypermultiplet
metric, forbidding any mass or $\Lambda$-dependent corrections to the
Higgs branch.  A similar argument also shows that the Coulomb branch
cannot receive any squark--vev dependent corrections, and that the
mixed branches have the structure of a direct product of a Higgs and
Coulomb branch.

This also implies that there are no non-trivial fixed points even at
strong coupling on a Higgs branch, except, perhaps, at points where it
joins a Coulomb branch.  Generic points on the Coulomb branch are free
$\CN{=}2$ $U(1)^{n_V}$ gauge theory in the IR.  Along certain
submanifolds of complex codimension one in moduli space the low-energy
theory contains in addition a massless charged hypermultiplet, giving
massless $\CN{=}2$ QED as the IR theory.  In terms of the original
gauge fields the light matter fields may be magnetic monopoles or dyons
of various charges.  Where these submanifolds intersect there will be
two or more massless charged hypermultiplets.  When the various
massless hypermultiplets at some point are all mutually local the
low-energy theory is simply $\CN{=}2$ electrodynamics with massless
matter, written in terms of the gauge fields to which the matter fields
couple locally.  The term ``mutually local'' as used here means simply
that there is an electric-magnetic duality transformation in the
low-energy $U(1)$ to a description of the physics in which no fields
carry magnetic charge.  The new phenomenon studied in \AD\ occurs when
the massless states at some point are mutually nonlocal, which, by the
preceding discussion, will be a nontrivial $\CN{=}2$ superconformal
field theory.

We can deform any fixed point on the Coulomb branch by vevs of
$\CN{=}2$ vector superfields to another point on the Coulomb branch, or
by relevant operators.  For simplicity, consider the case $n_V{=}1$, so
that at a generic point on the Coulomb branch we have a free Maxwell
theory described by a free $U(1)$ $\CN{=}2$ vector superfield $U$, with
$D(u) {=}1$.  The effect of adding a relevant operator appears in the
low energy theory as a variation of the prepotential.  An example is a
shift of the mass term for the underlying quarks in the microscopic QCD
Lagrangian.  The leading order operator at a given point on the moduli
space can be found by expanding the resulting variation of the
prepotential in $U$.  The constant term obviously does not contribute.
The linear term, $\int\! d^4\theta\, U$, is a total
derivative.\foot{$\int\! d^4\theta$ is an integral over half of
$\CN{=}2$ superspace.}  This follows by using the chirality
\chirality\ and null vector \nullvect\ conditions.  Therefore, the
leading effect of the mass term is to change the effective coupling
$\tau$ (the coefficient of $U^2$).  In $\CN{=}2$ QED ({\it i.e.} along
the codimension one submanifolds of the Coulomb branch where a charged
hypermultiplet becomes massless) a mass term can again be absorbed in a
shift of $U$, and again the term linear in $U$ does not have an effect.

At a non-trivial fixed point, on the other hand, $U$ is chiral, but no
longer satisfies \nullvect.  Expanding around $U{=}0$, the leading
effect of a mass term is then
		\eqn\veccoup{
\delta{\cal L} = \int\! d^4\theta\, mU,
		}
implying $D(m) + D(u) = 2$.

Alternatively, we can think of all the parameters as background gauge
fields or as weakly coupled propagating gauge fields.  This turns the
relevant operator deformations into deformations along the Coulomb
branch of these new gauge fields.  The masses can be thought of as the
background values of the scalar components of $\CN{=}2$ vector
multiplets coupling to the conserved flavor currents.  Since conserved
charges are dimensionless, their conserved currents have dimension
three.  These, in turn, couple to the vector field in the multiplet,
whose dimension is thus required to be one.  The $\CN{=}2$ algebra
relates this to the dimension of the scalar component, leading to the
requirement that the masses have dimension one.

This conclusion depended on the flavor symmetry being a global
symmetry, so we could think of it as being arbitrarily weakly gauged.
The conclusion is quite different if the global symmetry vanishes in
the conformal field theory.  This can happen in practice when a global
$U(1)$ symmetry looks at low energies like a global $U(1)$ gauge
transformation (like lepton number in QED, which coincides with the
global electric charge).  Then, at a non-trivial fixed point of the
effective $U(1)$ theory, the gauge charge and thus the global symmetry
will vanish, as we have seen above.  There may also be other mechanisms
by which a global symmetry of a microscopic Lagrangian can become
trivial at a conformal fixed point.

When the conserved currents do not act in the CFT ({\it i.e.}, their
charges vanish for all fields in the CFT) the superfield weakly gauging
the symmetry can couple to the CFT only through irrelevant operators.
For example, consider weakly gauging a $U(1)$ symmetry with gauge field
$V$ whose dimension is one.  Its leading coupling to the CFT is through
the irrelevant operator
		\eqn\irrcoup{
\Lambda^{-\delta}\int\! d^4\theta\, VU,
		}
where $D(u)= 1{+}\delta$.  Then, as we give an expectation value $\vev
v$ to the scalar component of $V$, the CFT is deformed by the relevant
or marginal operator
		\eqn\reldef{
\delta{\cal L} = {\vev v\over\Lambda^\delta}\int\! d^4\theta\, U,
		}
if $\delta{\le}1$.  Therefore, we identify the singlet mass parameter
$m = \vev v \Lambda^{-\delta}$, which has dimension $1{-}\delta$.  If
$D(u){>}2$ ($\delta{>}1$) the deformation \reldef\ is irrelevant.

This can be summarized as follows:  deforming the CFT by a gauge
superfield $V$ with dimension one leads to explicit breaking of scale
invariance, while deforming by a field $U$ with dimension larger than
one leads to spontaneously broken scale invariance.  The deformation
\reldef\ could arise with $V$ one of the elementary gauge fields in a
non-Abelian theory, as in the $SU(3)$ Yang-Mills example of \AD, or from
singlet mass terms for elementary hypermultiplets, as in the $SU(2)$
QCD examples to be discussed in section 3.

This picture can be generalized to include many $U$'s, in which case the
CFT has a set of relevant or marginal parameters $(u_i, m_i)$ satisfying
$1 < D(u) \le 2$ and $D(u_i) + D(m_i) = 2$, as well as parameters $m_A$
governing the coupling to any non-Abelian global symmetries.  Examples of
this sort occur in $n{\ge}4$ $SU(n)$ Yang-Mills \AD.

\newsec{Examples in SU(2)}

In this section we study the non-trivial fixed points occurring in
$\CN{=}2$ QCD with $SU(2)$ gauge group.  To see that such vacua might be
expected in this theory, note that they can arise at most in complex
codimension two in moduli space, explaining their absence in the $SU(2)$
Yang-Mills solution of \SWI, where the moduli space is the complex
$\tu$-plane.  When matter multiplets in the fundamental are included,
however, the bare masses appear as parameters in the theory.  By tuning
these parameters we can find values for which various subsets of the
singular points in the $\tu$-plane coincide.  If the colliding
singularities correspond to mutually nonlocal states, there will be a
new CFT.  We present a complete catalog of such points arising in
$\CN{=}2$ QCD with $SU(2)$ gauge group, some of which are manifestly
inequivalent to the CFT discovered in \AD.

At a point in parameter space where some singular points in the moduli
space coincide, there will be new, interacting physics if the massless
states associated to the two colliding singularities are mutually
nonlocal.  This can be determined by a local monodromy computation, but
we will propose a much simpler criterion.  As one varies the
parameters, if at some point in moduli space an enlarged set of
mutually local states become massless then the dimension of the Higgs
branch should increase for this special value of the parameters (as
happens, for example, when the bare masses are tuned to coincide for
two or more quarks).  Thus, if two singularities collide for some value
of the parameters and the Higgs branch does not change at this value,
the colliding points correspond to mutually nonlocal states.  This
Higgs branch criterion is effective because the structure of the Higgs
branch is determined classically due to the nonrenormalization theorem
discussed in the last section.

The Higgs branch criterion immediately suggests the following structure
of fixed points.  Choose the bare masses for all $N_f$ quarks to be the
same, $M$, giving an unbroken $U(N_f)$ global flavor symmetry, and an
$N_f{-}1$ dimensional Higgs branch.\foot{For $M{=}0$ the flavor
symmetry is enhanced to $SO(2N_f)$ and the Higgs branch is
correspondingly enlarged.} There are then three singular points in the
(finite) $\tu$-plane with the massless hypermultiplets at one
transforming in the fundamental representation of the flavor symmetry
and those at the other two invariant.  For large $M$ these are
interpreted as (one component of) the original quarks at $\tu {\sim}
M^2$, and the monopole and dyon states of the $SU(2)$ Yang-Mills theory
obtained after integrating out the massive quarks.  By tuning $M$ so
that a ``quark'' point coincides with one of the other two,
we can find new singularities.  The
discussion of the previous paragraph tells us that if the coincidence
occurs for nonzero $M$, so that the global symmetry and Higgs branch
are not modified from their form at generic $M$, the low-energy theory
should be an interacting CFT.  We will call such points $(N_f,1)$
points, indicating that $N_f$ mutually local states are massless
together with one state nonlocal with respect to them.

We thus predict the existence of these new CFT's.  In the rest of this
section we show that these points
exist for $N_f {\le} 4$, and that this list is comprehensive.  The
exact solution is written in terms of a cubic plane curve, describing a
torus as a branched cover of the $x$-sphere.  The coefficients of the
cubic polynomial are themselves polynomials in $\tu$ and the masses
$M_i$, $i {=} 1, \ldots, N_f$. Singular points in the
$\tu$-plane correspond to degenerations of the torus, or equivalently to
the coincidence of roots of the polynomial.  The interacting points in
codimension two will arise when all three of the roots at finite $x$
coincide.  Because three branch points coincide at these points, two
intersecting cycles on the torus are going to zero, and thus the light
hypermultiplets near these points are mutually nonlocal.  In the
vicinity of such a point, we can write the curve in terms of local
coordinates $u$, $m$, $\til x$ as $y^2 = \til x^3 {-} f(u,m)\til x {-}
g(u,m)$, where $f,g$ are polynomials such that $f(0,0)=g(0,0)=0$.
Keeping only the lowest-order terms in $f$, $g$ gives a
quasihomogeneous polynomial, leading to an assignment of $R$-charges
and hence to a prediction of the anomalous dimensions of some of the
chiral operators in the CFT.

\subsec{$N_f=1$}

The one-flavor curve\foot{We set $\Lambda^{4-N_f} {=} 8$ by a choice of
units.  The appropriate powers of $\Lambda$ can easily be reinstated
using the $R$-symmetry charges.} $y^2 = x^2(x{-}\tu) + 2Mx - 1$ is
singular at the zeros of the discriminant $\Delta_1 = -4\tu^3 {+} 4\tu^2M^2
{+} 36\tu M {-} 32M^3 {-} 27$.  Special points can arise when two or more
of the zeros of $\Delta_1$ coincide. This must necessarily happen,
since the discriminant of $\Delta_1$ considered as a cubic in $\tu$ is
holomorphic in $M$ and hence vanishes somewhere.  Indeed the special
point occurs at $M {=} {3\over2}\omega$, $\tu {=} 3\omega^{-1}$, for
which three branch points coincide at $x {=} \omega^{-1}$, where
$\omega^3 {=} 1$.\foot{The bare mass breaks a $\bZ_3$ symmetry acting
on the $\tu$-plane giving three singular values of $m$ with
identical structure.  In terms of the identification of the massless
states at large $M$, these three points correspond
to the quark, monopole, and dyon becoming massless in pairs.}
Expanding about this point (taking $\omega{=}1$) and rewriting the
curve in terms of the shifted variables $M = {3\over2} {+} m$, $\tu = 3
{+} 2m {+} u$, and $x = {1\over3}\tu {+} \til x$, we have
		\eqn\yonexpi{
y^2 = \til x^3 - 2(m+u)\til x-(u + \textstyle{4\over3}m^2),
		}
where we have dropped higher-order terms in $m$ and $u$ which are
necessarily smaller than those shown as $m,u{\to}0$.  Furthermore, in
this limit we must assign relative scaling dimensions $D(x):D(m):D(u) =
1:2:3$ in order to see the cubic singularity.  Then the $ux$ and $m^2$
terms are negligible near the CFT point, and we find the curve
		\eqn\yonexp{
y^2 = \til x^3 - 2m\til x-u .
		}
We discuss how the apparent scaling dimensions of the coefficients
of the curve translate into scaling dimensions of operators in the CFT
at the end of this section.

The curve \yonexp\ is identical to the expansion of the $SU(3)$
Yang-Mills curve around the point discovered in \AD.  Also, neither
theory has a Higgs branch or a flavor symmetry.  One difference is that
here $m$ appears as a parameter whereas in \AD\ its place was taken by
the background value of a field; as discussed above, this difference is
not essential.  Indeed, it was shown in \DW\ that these two CFT's are
the same by realizing them as different limits related by S-duality of
the $\CN{=}2$ $SU(3)$ theory with one adjoint flavor.  We refer to this
class of special point as $(1,1)$ indicating that it arises when the
singularities corresponding to two, mutually nonlocal massless states
coincide.

\subsec{$N_f=2$}

For two flavors the curve is $y^2 = (x^2{-}1) (x{-}\tu) + 2M_1M_2x -
(M_1^2{+}M_2^2)$.  The four values of $\tu$ at which this is singular are
given by the vanishing of its discriminant $\Delta_2$.  To determine
when these zeros collide we can study the discriminant in $\tu$ of
$\Delta_2$, which has a factor corresponding to the lines $M_1 = \pm
M_2$ of coincidence of the two electron points when their bare masses
are degenerate, and another factor $\Delta$ along which there are
additional coincidences.  We can obtain $\Delta$ more directly by
requiring that the two-flavor curve be an exact cube $y^2 = (x {-}
{1\over3}\tu)^3$, yielding $\tu^2 = 6M_1M_2 {-} 3$ and $\tu^3 {+} 27\tu
= M_1^2 {+} M_2^2$.  Eliminating $\tu$ gives a locus $\Delta(M_i) = 0$ of
singularities in $M$-space.

Taking the scaling limit at large $M_2$ (and vanishing $\Lambda$)
leads to the $(1,1)$ point found above.  By continuity, we expect a
line of such points along any branch of $\Delta$, extending from the
large-$M$ region until a new singularity appears.  There are two
possibilities here:  either the low-energy theory remains constant as
we move $M$ along this curve and tune $\tu$ to the singular value, or
they are connected by an exactly marginal operator.  The former seems
more likely, since there is no continuous parameter in the $(1,1)$
curve \yonexp\ which could correspond to the parameter of the marginal
deformation.  In particular, the coefficients in \yonexp\ can be
absorbed in simple redefinitions (rescalings) of $m$ and $u$.

Additional singularities occur when $\Delta$ intersects one of the
lines $M_1 = \pm M_2$, or when it is itself singular.  There are
singularities of $\Delta$ at the four solutions of $M_1^2 {+} M_2^2 =
0$, $M_1 M_2 = -4$.  These points contain no new physics; the apparent
singularity is due to the existence of two different values of $\tu$ for
the same $M_i$ at which the theory is singular.  The other
singularities of $\Delta$ in $M$-space occur only at $M_1=\pm M_2$, for
the values $M_1^2 = \pm 2$.  These points {\it do} correspond to new
physics since a $U(2)$ flavor symmetry is unbroken by the bare masses,
and the Higgs branch of the theory (which is constant along the line
$M_1 = \pm M_2$) does not change at this point.  This indicates a
nonlocal collision, and also shows that this CFT cannot be equivalent
to the $(1,1)$ theory.

To study this point, rewrite the two-flavor curve in terms of the
flavor invariants $M \equiv {1\over2}\sum_i M_i$ and $C_2 \equiv
\sum_i(M_i-M)^2$, in terms of which the point in question is $M {=}
\sqrt2$, $C_2 {=} 0$.  Expand about this point, defining the shifted
variables $M= \sqrt2 {+} m$, $\tu = 3 {+} 2\sqrt2 m {-} {1\over3}m^2
{+} u$, and $x = {1\over3}u {+} \til x$, to give the curve
		\eqn\ytwoexp{
y^2 = \til x^3 - 2u\til x - {4\sqrt2\over3}mu + {16\sqrt2\over27}m^3 -
2C_2 .
		}
The structure of the theory around this point can be read from
\ytwoexp.  The dimensions of the relevant couplings are in the ratios
$D(m):D(u):D(C_2) = 1:2:3$.  In the vicinity of the special point
there are distinguished deformations.  Varying the singlet mass $m$ with
$C_2 {=}0$ preserves the
$U(2)$ global symmetry; the singularity in the $u$-plane splits, one
point corresponding to two massless states in the {\bf 2} and the other
to a singlet. The Higgs branch is unchanged under this deformation.
Further, one expects to find $(1,1)$ points in the vicinity of the
$(2,1)$ point.  Indeed, \ytwoexp\ factors as a cubic for $4m^3 {=}
9\sqrt2 C_2$, $u {=} -{1\over3}m^2$; this is the local form of $\Delta$
about this point.

As an illustration of the Higgs branch criterion in action, consider
the case when the two quarks have a common mass $M$.  For large $M$
there will be a 2-quark point at $\tu {\sim} M^2$, and a monopole and a
dyon point at $\tu \sim \pm1$.  There is a Higgs branch isomorphic to
$\bC^2/\bZ_2$ attached to the 2-quark point, and the low-energy theory
has an unbroken global $U(2)$ flavor symmetry;  the monopole and dyon
points have no Higgs branches and only a $U(1)$ baryon number.
Classically, decreasing $M$ gives no qualitatively new physics until $M
{=} 0$, at which point the global symmetry is enhanced to $SO(4)$ and
two Higgs branches appear, both isomorphic to $\bC^2/\bZ_2$.  This must
also be true quantum-mechanically by the nonrenormalization theorem.
The Higgs branch criterion implies that collisions of monopole points
at $M {=} 0$ where the dimension of the Higgs branch changes will
correspond to a vacuum with only mutually local states.

On the other hand, from the exact solution, we expect three
singularities as we decrease $M$:  two where the 2-quark point
coincides with a monopole or dyon point, and one where the monopole and
dyon points meet.  Naively, all these singularities involve massless
mutually non-local hypermultiplets.  As shown above, the first two of
these meetings take place at $M = \pm\sqrt 2$ and $\tu{=}3$, and indeed
involve mutually nonlocal states since three branch points coincide at
that point.  The other singularity occurs at $M{=}0$, where the curve
becomes $y^2 = (x^2{-}1) (x{-}\tu)$.  Here there are two singularities in
the $\tu$-plane, corresponding to massless monopole states in one spinor
of $SO(4)$ at $\tu{=}1$, and massless dyons in the other spinor of
$SO(4)$ at $\tu{=}-1$ \SWII.  We see directly from the curve that the
vacua involve only mutually local massless particles.  The naive
picture that this vacuum arose from the collision of a monopole and
dyon point, and therefore should have involved mutually nonlocal
massless states, is incorrect because it did not take into account the
possibility that as $M{\to}0$ the monodromies around the monopole or
dyon point are conjugated by the 2-quark point monodromy as that point
moves on the $\tu$-plane.  Indeed, one can check that this is what
happens: the 2-quark singularity passes between the monopole and dyon
points as $M{\to}0$.

\subsec{$N_f=3$}

The curve here is $y^2 = x^2(x{-}\tu)-(x{-}\tu)^2-A(x{-}\tu)+2Bx-C$,
where $A = M_1^2 {+} M_2^2 {+} M_3^2$, $B = M_1M_2M_3$, and $C =
M_1^2M_2^2 {+} M_2^2M_3^2 {+} M_1^2M_3^2$.  To find $(1,1)$ points set
$y^2 = (x {-} {1\over3}\tu {-} {1\over3})^3$ and eliminate $\tu$ to find a
surface $\Delta(M_i)=0$.
%		\eqn\delthree{\eqalign{
%\Delta &= 1-12A-270C+168B+48A^2+216AC+168AB-27C^2-216CB-\cr
%&\quad636B^2-64A^3+ 96A^2B+60AB^2+8B^3 .\cr
%		}}
Along this surface in $M$-space, the generic singular form of the
three-flavor curve describes $(1,1)$ vacua.  The surface is singular
along a submanifold where two of the bare masses coincide; for these
masses there are $(2,1)$ points with the appropriate Higgs branch and
$U(2)$ global symmetry.  This submanifold is itself singular when all
of the masses are equal, leading to a $(3,1)$ point occurring at $M_1
{=} M_2 {=} M_3 {=} 1$ with a $U(3)$ flavor symmetry.  A detailed study
of $\Delta$ shows that there are no other singularities, thus we expect
no other nontrivial CFT's to arise.

The nonzero bare masses at the $(3,1)$ point mean the Higgs branch is
not changed and hence there is a new non-trivial CFT here.  To study
this point rewrite the three-flavor curve in terms of the invariants $M
\equiv {1\over3}\sum_i M_i$, $C_2 \equiv \sum_i (M_i {-} M)^2$, and
$C_3 \equiv \sum_i (M_i {-} M)^3$, in terms of which the point in
question is $M{=}1$, $C_2{=}C_3{=}0$.  We expand about this point and
shift $M = 1 {+} m$, $\tu = 2 {+} 3m {+} m^2 {-} {5\over6}m^3 {+} u$, and
$x = {1\over3} {+} {1\over3}u {+} \til x$.  Writing the curve in terms
of the shifted variables and dropping higher orders we have
		\eqn\ythreexp{
y^2 = \til x^3 - 2(mu+C_2)\til x - u^2 - {m^3u\over3} + {m^6\over108} -
{2m^2C_2\over3} + {8C_3\over3} .
		}
The relevant couplings now satisfy $D(m):D(u):D(C_2):D(C_3) =
1:3:4:6$.  There is a surface of $(1,1)$ points coinciding with the
local form of $\Delta$.

\subsec{$N_f\ge $4}

For $N_f{=}4$, the bare masses $M_i$ break the exact scale symmetry of
the massless theory \SWII.  This means that an overall scaling of the
masses is not a true parameter of the theory, so that one cannot expect
a $(4,1)$ point to appear at codimension four.  On the other hand,
scale invariance means the classical coupling $\tau$ is a parameter in
the theory; the coefficients of the polynomial are modular functions of
$\tau$. One expects to find (though we did not check this) a
hypersurface in $M$-space, varying with $\tau$, along which there are
$(1,1)$ points, singularities on this corresponding to $(2,1)$ and then
$(3,1)$ points.  We did look for the possible occurrence of a $(4,1)$
point at a particular value of $\tau$, and found that this does not
happen. Of course, we already know of a new conformal fixed point that
occurs for $N_f{=}4$, namely the non-Abelian Coulomb point at
$M_i{=}\tu{=}0$.  By a series of shifts and rescalings, the curve near
this point can be put into the form
		\eqn\yfourexp{
y^2 = \til x^3 + (u^2 + uC_2 + C_2^2 + C_4 + C_4') \til x +(u^3 + u^2
C_2 + \ldots + C_6) ,
		}
where we have suppressed the coefficients, which are various modular
functions of the coupling $\tau$, and where the $C_i$ are the
mass-invariants of the $SO(8)$ flavor group: $C_2 \equiv \sum_i M_i^2$,
$C_4 \equiv \sum_{i<j} M_i^2 M_j^2$, $C_4' \equiv \prod_i M_i$, and
$C_6 \equiv \sum_{i<j<k} M_i^2 M_j^2 M_k^2$.  In this case the
$R$-charges of the relevant and marginal couplings $u$, $\tau$, $C_2$,
$C_4$, $C_4'$, and $C_6$, are just proportional to their classical
$R$-charges.

For $N_f {>} 4$ the $SU(2)$ theory is no longer asymptotically free,
and instead has a free non-Abelian Coulomb phase at $\tu{=}0$.  The
Coulomb branch can be described in the vicinity of this point
(by breaking to $SU(2)$ from an asymptotically-free $SU(N_c)$ gauge
theory with $N_f$ flavors, as described in \APS) by the curve
                \eqn\IRfree{
y^2 = (x^2-\tu)^2-\Lambda^{4-N_f}\prod_{i=1}^{N_f}(x-M_i) .
                }
This description is only reliable for $x$, $\tu$, and $M_i \ll \Lambda$;
values of the parameters and moduli outside this region probe physics
that depends on the UV regulator.  For small but non-zero $M_i {\sim}
M$, the monopole and dyon singularities of the effective $SU(2)$
Yang-Mills theory at energies less than $M$ occur at $\tu^2 \sim
\Lambda^4 (M/\Lambda)^{N_f}$.  Thus the quark singularities at $\tu
{\sim} M^2$ can only approach the monopole and dyon singularities for
$M {\sim} \Lambda$ where the curve \IRfree\ is no longer valid.  We
conclude that there are no new interacting vacua in $\CN{=}2$ $SU(2)$
QCD with $N_f{>}4$.

\subsec{The $R$-Symmetry}

The quasihomogeneous polynomials (3.2-5) determine the $R$-charges of
the various couplings, and hence their dimensions, up to an overall
scaling.  A constraint which fixes the overall normalization of the
$R$-charges follows from the requirement that the K\"ahler potential $K
= {\rm Im}(A_D\bar A)$ have dimension two.  This, in turn, means that
$a$ has dimension one.  Using the representation of $a$ as a contour
integral on the torus specified by the cubic curve, we have $a\sim
(u/y)d\til x$, which then leads to a normalization of the dimensions.

{\par\begingroup\parindent=0pt\leftskip=1cm\rightskip=1cm\baselineskip=11pt
 \midinsert\centerline{\vbox{\offinterlineskip
\halign{&\vrule width 1.2pt#&\strut\quad\hfil#\hfil\quad&
 \vrule#&\strut\quad\hfil#\hfil\quad&\vrule#&\strut\quad\hfil#\hfil\quad&
 \vrule#&\strut\quad\hfil#\hfil\quad&\vrule#&\strut\quad\hfil#\hfil\quad&
 \vrule#&\strut\quad\hfil#\hfil\quad&\vrule#&\strut\quad\hfil#\hfil\quad&
 \vrule width 1.2pt#\cr
    \noalign{\hrule height 1.2pt}
    height3pt&\omit&&\omit&&\omit&&\omit&&\omit&&\omit&&\omit&\cr
    &$N_f$&&$y$&&$\til x$&&$u$&&$m\ (\tau)$&&$m_A$&&$a$&\cr
    height3pt&\omit&&\omit&&\omit&&\omit&&\omit&&\omit&&\omit&\cr
    \noalign{\hrule}
    height3pt&\omit&&\omit&&\omit&&\omit&&\omit&&\omit&&\omit&\cr
    \noalign{\hrule}
    height3pt&\omit&&\omit&&\omit&&\omit&&\omit&&\omit&&\omit&\cr
    &1&&3/5&&2/5&&6/5&&4/5&&&&1&\cr
    height3pt&\omit&&\omit&&\omit&&\omit&&\omit&&\omit&&\omit&\cr
    \noalign{\hrule}
    height3pt&\omit&&\omit&&\omit&&\omit&&\omit&&\omit&&\omit&\cr
    &2&&1&&2/3&&4/3&&2/3&&1&&1&\cr
    height3pt&\omit&&\omit&&\omit&&\omit&&\omit&&\omit&&\omit&\cr
    \noalign{\hrule}
    height3pt&\omit&&\omit&&\omit&&\omit&&\omit&&\omit&&\omit&\cr
    &3&&3/2&&1&&3/2&&1/2&&1&&1&\cr
    height3pt&\omit&&\omit&&\omit&&\omit&&\omit&&\omit&&\omit&\cr
    \noalign{\hrule}
    height3pt&\omit&&\omit&&\omit&&\omit&&\omit&&\omit&&\omit&\cr
    &4&&3&&2&&2&&0&&1&&1&\cr
    height3pt&\omit&&\omit&&\omit&&\omit&&\omit&&\omit&&\omit&\cr
    \noalign{\hrule height 1.2pt}}}}
 \vskip 12pt {\bf Table 1:} Scaling weights of the couplings at $(N_f,1)$
    points for various $N_f$.  The adjoint masses are defined by $m_A =
    (C_j)^{1/j}$.  For $N_f{=}4$, $\tau$ plays the role of the scalar mass
    $m$.  \par\endinsert\endgroup\par}

We collect the data on the four CFT's, normalized in this way, in Table
1.  The non-integral spectra of $R$-charges demonstrate that the
$R$-symmetry in the $\CN{=}2$ superconformal algebra is not a symmetry
of the classical Lagrangian for $N_f {\neq} 4$.  In addition, the
spectrum of dimensions shows that these are not free field theories.
The first example of an $\CN{=}2$ SCFT in four dimensions exhibiting this
property is the $(1,1)$ theory found in $SU(3)$ Yang-Mills theory \AD.
A number of properties are immediately clear from the table:
\item{1.} The dimension of $m_A = (C_j)^{1/j}$ is one (in the cases in
which it is defined).
\item{2.} The dimension of $u$ satisfies $1 < D(u) \le 2$.
\item{3.} The dimensions of $u$ and $m$ satisfy $D(u) {+} D(m) = 2$.

\noindent The explanation of properties 1--3 in terms of $\CN{=}2$ CFT
was given in section 2.  Property 1 follows from the fact that
the adjoint mass couples to conserved flavor currents which
act in the CFT.
Property 2 reflects the fact that these are interacting CFT's with
$u$ (the vev of) a relevant operator.  Property 3 follows from the form
of the relevant coupling \veccoup.

We can give no explanation of the spectrum of dimensions in Table 1
{}from first principles; however, assuming a cubic form for the
singularity, one can recover the information in Table 1.  In
particular, assume that the curve around the CFT point is given by $y^2
=x^3 {-}fx {-}g$ where $f$, $g$ are polynomials in $u$, $m$, and $m_A$,
which correspond to relevant operators.  Normalizing the dimensions by
demanding that $a \sim (u/y)dx$ have dimension one, and using
properties 1--3 above, gives $D(m_A) {=} 1$, $D(m) {=} 2 {-} D(u)$,
$D(f) {=} 4D(u) {-}4$, $D(g) {=} 6D(u) {-} 6$, and $D(x) {=} 2D(u) {-}
2$.  Either $f$ or $g$ must include a term $u^\alpha$ for $\alpha {=}
1, 2, \ldots$, since otherwise the curve would be singular for all $u$
when $m {=} 0$, implying that $u$ is not, in fact, a relevant operator
as we had assumed.  Assuming first $g {\sim} u^\alpha$ one finds that
only the values $\alpha {=} 1,2,3$ are compatible with with the above
constraints.  For $\alpha {=} 1$ one finds the $(1,1)$ curve; for
$\alpha {=} 2$ the $(3,1)$ curve; and for $\alpha {=} 3$ the $(4,1)$
curve.  On the other hand, assuming $f \sim u^\beta$ gives the $(2,1)$
curve for $\beta {=} 1$; coincides with the $\alpha {=} 3$ case when
$\beta {=} 2$; and no other $\beta$ are allowed.  We thus recover
precisely the singular curves found above.  Furthermore, many (though
not all) of the coefficients of the various terms in the curves can be
taken to 1 by appropriate rescalings and shifts of $u$, $m$, and
$m_A$.

One could imagine developing in this way an algebraic classification of
four-dimensional $\CN{=}2$ CFT's.

\bigskip
\centerline{{\bf Acknowledgements}}

It is a pleasure to thank T. Banks and M. Douglas for helpful
discussions and comments.  This work was supported in part by grants
DOE DE-FG05-90ER40559 and NSF PHY92-45317.

\listrefs
\end